\documentclass[12pt,english,preprint,preprintnumbers]{revtex4}
\usepackage{mathpazo}
\usepackage[T1]{fontenc}
\usepackage[latin9]{inputenc}
\usepackage{color}
\usepackage{amsmath}
\usepackage{amssymb}
\usepackage{cancel}

\makeatletter

\providecommand{\tabularnewline}{\\}

\@ifundefined{textcolor}{}
{%
 \definecolor{BLACK}{gray}{0}
 \definecolor{WHITE}{gray}{1}
 \definecolor{RED}{rgb}{1,0,0}
 \definecolor{GREEN}{rgb}{0,1,0}
 \definecolor{BLUE}{rgb}{0,0,1}
 \definecolor{CYAN}{cmyk}{1,0,0,0}
 \definecolor{MAGENTA}{cmyk}{0,1,0,0}
 \definecolor{YELLOW}{cmyk}{0,0,1,0}
}

\usepackage{babel}

\usepackage{amsfonts}
\usepackage{epsfig}
\usepackage{bm}

\usepackage{color}

\definecolor{darkpastelpurple}{rgb}{0.59, 0.44, 0.84}
\definecolor{frenchlilac}{rgb}{0.53, 0.38, 0.56}
\definecolor{violet}{rgb}{0.56, 0.0, 1.0}

\usepackage{graphicx}

\usepackage{verbatim}
\usepackage{ae,aecompl}
\usepackage{amsmath,amssymb,mathrsfs}

\providecommand{\ZZ}{\mathbb{Z}}

\usepackage{capt-of}


\usepackage{adjustbox}

\usepackage{etoolbox}
\makeatletter
\patchcmd{\maketitle}
 {\def\@makefnmark}
 {\def\@makefnmark{}\def\useless@macro}
 {}{}
 \makeatother

\makeatother

\usepackage{babel}
\begin{document}

	\author{\normalsize { \bf B. L. Sánchez--Vega$^{1,*}$}, 
 		{\bf E. R. Schmitz $^{2,\dagger}$}
 		and {\bf J. C. Montero $^{1,\star}$}\\ 
 		{\footnotesize \it $^{1}$ Universidade Estadual Paulista (Unesp), Instituto de Física Teórica (IFT), São Paulo \\
 		R. Dr. Bento Teobaldo Ferraz 271, Barra Funda, São Paulo - SP, 01140-070, Brazil }\\
 		{\footnotesize \it $^{2}$ Bethe Center for Theoretical Physics and Physikalisches Institut, Universität Bonn, Nussallee 12, D-53115 Bonn, Germany.}}
 		
 		\title{New constraints on the 3-3-1 model with right-handed neutrinos\footnotetext{$^*$brucesan@ift.unesp.br\\
 		 		$^\dagger$ernany@th.physik.uni-bonn.de\\
 		 		$^\star$montero@ift.unesp.br}}







\begin{abstract}
\textcolor{black}{In the framework of a }$3-3-1$\textcolor{black}{{}
model with right-handed neutrinos and three scalar triplets we consider
different spontaneous symmetry breaking patterns seeking for a non-linear
realization of accidental symmetries of the model, which will produce
physical Nambu-Goldstone (NG) bosons in the neutral scalar spectrum.
We make a detailed study of the safety of the model concerning the
NG boson emission in energy loss processes which could affect the
standard evolution of astrophysical objects. We consider the model
with a }$\ZZ_2$ symmetry, conventionally used in the literature,
finding that in all of the symmetry breaking patterns the model is
excluded. Additionally, looking for solutions for that problem, we
introduce\textcolor{black}{{} soft }$\ZZ_2$\textcolor{black}{-breaking
terms in the scalar potential} in order to remove the extra accidental
symmetries and at the same time maintain the model as simple as possible.
We find that there is only one \textcolor{black}{soft }$\ZZ_2$\textcolor{black}{-violating
term that can get rid of the problematic NG bosons. }
\end{abstract}

\maketitle

\section{Introduction\label{introduction}}

Recently experiments have reached the capability of exploring the
TeV energy scale and the Standard Model (SM) still has an impressive
accordance with data. However, there are at least two important aspects
which put in evidence the incompleteness of the SM and, hence, the
need for new physics. Namely they are i) the non-zero neutrino masses,
which allows for neutrino flavor oscillation during space propagation,
and ii) the lack of a consistent candidate for dark matter (DM), assuming
that DM is a manifestation of an unknown particle.

In this way, models based on the $\textrm{SU}\left(3\right)_{C}\otimes\textrm{SU}\left(3\right)_{L}\otimes\textrm{U}\left(1\right)_{N}$
gauge symmetry (the so called $3-3-1$ models, for shortness) \cite{Pleitez_etal_1992,Frampton_1992,Montero:1993,Z2_331_3},
are interesting extensions of the SM. Since in these models the electroweak
interaction is supposed to be invariant under transformations of a
larger gauge group, the matter content can be chosen to accommodate
new appropriate degrees of freedom in order to implement phenomenologically
attractive features, as generation of neutrino masses, for instance. 

The quark sector will also be enlarged and new quarks will be present,
with the possibility of possessing exotic electric charges. Chiral
anomaly cancellation is assured provided we have the same number of
triplets and anti-triplets, including color counting. If we assume
that there is a symmetry between leptons and quarks in such a way
that the number of families of leptons is equal to the number of families
of quarks, say $N_f$, then, we must have $N_{\textrm{anti}}=2N_f/3$
quark families transforming under the $\bar{\bf{3}}$ representation
of the SU$(3)_L$ group, and the other $N_f - N_{\textrm{anti}}$
families transforming under the $\bf{3}$ one. It means that the number
of families must be three or a multiple of three. As a consequence,
differently from the SM, the model is anomaly free only when the total
number of families is considered. 

Moreover, if we bring in the QCD asymptotic freedom, the number of
families must be just 3. The renormalization group $\beta$ function,
which gives the behavior of the strong coupling with the transferred
momentum, can be computed in perturbation theory and at one-loop level
its sign is governed by the factor $-(11-\frac{2}{3}n_q)$, where
$n_{q}$ is the number of quark flavors (which is $6$ in the SM).
To keep the negative sign, the only possibility compatible with asymptotic
freedom, $n_q$ must be $\leq 16$. In $3-3-1$ models with $N_f=3$
we have $9$ quark flavors and the sign of the $\beta$ function remains
correct. However, for $N_f=6$ there are $18$ quark flavors and the
$\beta$ function gets the wrong sign so that the number of families
must be $3$. It is interesting to note that the anomaly cancellation
relates the number of families to the number of colors. In this sense
the $3-3-1$ models shed some light on the family replication problem.

Besides this particular feature, $3-3-1$ models are promising alternatives
to the SM for they present a variety of interesting properties. Among
them we can mention: i) in the model described in Ref. \cite{Pleitez_etal_1992}
we find that the U$(1)_{N}$ and the SU$(3)_{L}$ coupling constants,
$g_{N}$ and $g_{L}$, respectively, obey the relation $t^{2}=(g_{N}/g_{L})^{2}=\sin^{2}\theta_{W}/(1-4\sin^{2}\theta_{W})$.
It means that there is a Landau-like pole at an ${\cal O}(\mathrm{TeV})$
energy scale $\mu$ such that $\sin^{2}\theta_{W}(\mu)=1/4$ \cite{Dias2005},
and hence it explains why $\sin^{2}\theta_{W}(\mu)<1/4$ is observed;
ii) the electric charge quantization is independent of the nature
of neutrinos, i.e., regardless if they are Majorana or Dirac fermions
\cite{Pires:1998}; iii) the Peccei-Quinn symmetry, needed to solve
the strong CP problem, is almost automatic in these models \cite{Dias:2004,Bruce_Juan_331_PQ}.

For each $3-3-1$ model, depending on the matter content accommodated
in the SU$(3)_L$ triplets and singlets, an appropriate scalar sector
has to be introduced. As usual, the Yukawa terms are responsible for
generating mass to the matter fields and also for matter fields-(pseudo-)scalar
interactions. Therefore, the $3-3-1$ scalar sector is richer than
that of the SM, and this fact can be explored to give explanations
to some phenomenological aspects that do not have a consistent answer
or are out of the scope of the SM framework. Among others, some aspects
closely related to the scalar sector and the Yukawa interactions are
i) a mechanism for generating tiny neutrino masses \cite{Mizukoshi_Pires_Queiroz_daSilva_331,neutrino_masses,neutrino_masses_2},
ii) a natural explanation for the fermion mass hierarchy, and iii)
a consistent DM candidate \cite{Mizukoshi_Pires_Queiroz_daSilva_331,331_Farinaldo,331_darkmatter_1,331_darkmatter_2}. 

In order to achieve these goals, and recover the low energy physics,
spontaneous symmetry breaking (SSB) must occur. For each linearly
independent broken generator, there will be a Nambu-Goldstone (NG)
boson, if the number of broken symmetries exceeds the number of massive
gauge bosons, this implies physical massless NG bosons which are potentially
dangerous since they couple to fermions and could in principle escape
from star nuclei carrying out energy, thus modifying the standard
evolution of these objects. Their interactions with nucleons \cite{Tabulation_astrophysical_1987,astrophysical_2008,astro_1989,astro_1990,astrophysical_1996,neutron_stars_old1,neutron_stars_old2}
and electrons \cite{Tabulation_astrophysical_1987,RedGiant_Raffelt_1995,eeJ_old1,eeJ_old2,eeJ_old3},
are parameterized by $g_{nnJ}$ and $g_{e\bar{e}J}$, respectively,
(where $J$ means the physical NG boson), are bounded by the standard
evolution of neutron stars and supernovae; red giants, super giants,
sun, white dwarfs and neutron star crusts, and can constrain the parameters
of a given considered scenario or even rule it out. This is the main
goal of this work.

This work is organized as follows. In the Sec. \ref{sec:The-Model},
we present the general features of the $3-3-1$ model, including its
matter content, Yukawa and scalar potential. In the Sec. \ref{sec:Model-with-Z2},
we consider the model with a $\mathbb{Z}_{2}$ symmetry widely used,
and present the consequences of this choice, such as the number and
form of physical NG bosons for each display of vacuum expectation
values (VEVs), and a discussion of the constraints from $g_{e\bar{e}J}$
and $Z$ invisible decay on them. Finally, in the Sec. \ref{sec:Model-with-soft},
we add soft $\ZZ_2$-breaking terms to the previous scalar potential
and analyze their consequences using as a guidance the constraints
from the $g_{e\bar{e}J}$ and $g_{nnJ}$ couplings, and also the $W$
mass and the $\rho$ parameter values; a discussion of the constraints
at a symmetry point of view is also presented. Sec. \ref{conclusions}
is devoted to our conclusions.

\section{\label{sec:The-Model}The Model\label{model}}

The $3-3-1$ model with right-handed neutrinos considered in this
paper was proposed in Ref. \cite{neutrino_masses} and it has been
subsequently considered in Refs. \cite{Pires_daSilva_331,Dias_Pires_daSilva_331,Z2_331_2,Dong:2007,Mizukoshi_Pires_Queiroz_daSilva_331,Bruce_Juan_331_PQ,331_Farinaldo,Montero:2015},
where different aspects of this model were studied. This model shares
appealing features with other versions of $3-3-1$ models \cite{Pleitez_etal_1992,Frampton_1992,Clavelli_etal_1974,Lee_Weinberg_1977,Lee_Shrock_1978,Singer_1979,Singer_etal_1980,Montero:1993}.
It offers for example an explanation to the number of the fermion
families and makes easily the implementation of the Peccei-Quinn mechanism
\cite{Bruce_Juan_331_PQ}. Furthermore, right-handed neutrinos are
in the same multiplet of the SM leptons, which allows terms of mass
for the neutrinos at tree level, although the smallness of the those
masses remains unexplained.

Generally speaking, this model is based on the gauge symmetry group
$\textrm{SU}\left(3\right)_{C}\otimes\textrm{SU}\left(3\right)_{L}\otimes\textrm{U}\left(1\right)_{N}$,
where $C$ stands for color and $L$ for left chirality, as in the
SM; and $N$ stands for a new charge different than the SM hypercharge
$Y$. The $N$ values are assigned in order to obtain the SM hypercharge
$Y=2N\,\mathbf{1}_{3\times3}-\frac{1}{\sqrt{3}}T_{8}$, after the
first spontaneous symmetry breaking, and as a consequence the electric
charge $Q=T_{3}+\frac{Y}{2}\,\mathbf{1}_{3\times3}$, where $T_{3}$,
$T_{8}$ are the diagonal SU$\left(3\right)_{L}$ generators, whereas
$T_{9}\left(=\frac{N}{2}\,\mathbf{1}_{3\times3}\right)$ is the generator
of the U$\left(1\right)_{N}$ group. Symmetry breaking and fermion
masses are achieved with at least three SU$\left(3\right)_{L}$ triplets,
$\eta,\,\rho,\,\chi$, as shown in Ref. \cite{Montero:2015}. These
triplets are in the $\left(1,\:\mathbf{3},\,-1/3\right)$, $\left(1,\,\mathbf{3},\,2/3\right)$
and $\left(1,\:\mathbf{3},\,-1/3\right)$ representations of the $\textrm{SU}\left(3\right)_{C}\otimes\textrm{SU}\left(3\right)_{L}\otimes\textrm{U}\left(1\right)_{N}$
symmetry groups, respectively. In more detail, the scalar triplets
are expressed by \[\eta=
\left(
\begin{array}{ccc}
\eta^{0}_{1},\,\eta^{-}_{2},\,\eta^{0}_{3}\end{array}\right)^\textrm{T}, 
\quad
\rho=\left(\begin{array}{ccc} \rho^{+}_{1},\, \rho^{0}_{2},\, \rho^{+}_{3} \end{array}\right)^\textrm{T},
\quad 
\chi=\left(\begin{array}{c} \chi^{0}_{1},\, \chi^{-}_{2},\, \chi^{0}_{3} \end{array}\right)^\textrm{T}. \]

The fermionic content of the model is richer than the SM because the
fields are embedded into non-trivial representations of a larger group,
SU$\left(3\right)_{L}$. The left-handed fields belong to the following
representations
\begin{align}
\textrm{\,\ Leptons: }f_{aL} & =\left(\begin{array}{ccc}
\nu_{a} & e_{a} & N_{a}^{c}\end{array}\right)_{L}^{\textrm{T}}\sim\left(1,\,\mathbf{3},\,-1/3\right),\nonumber \\
\textrm{Quarks: }Q_{L} & =\left(\begin{array}{ccc}
u_{1} & \,d_{1} & \,u{}_{4}\end{array}\right)_{L}^{\textrm{T}}\sim\left(\mathbf{3},\,\mathbf{3},\,1/3\right),\nonumber \\
Q_{bL} & =\left(\begin{array}{ccc}
d_{b} & u_{b} & d_{b+2}\end{array}\right)_{L}^{\textrm{T}}\,\,\:\sim\left(\mathbf{3},\,\bar{\mathbf{3}},\,0\right),
\end{align}
where $a=1,\,2,\,3$ and $b=2,\,3$; and ``$\sim$'' means the transformation
properties under the local symmetry group. Notice that $N_{a}$ stands
for the right-handed neutrinos. Additionally, in the right-handed
field sector we have
\begin{align}
\textrm{Leptons: }e_{aR} & \sim\left(1,\,1,\,-1\right),\\
\textrm{Quarks:\,\ }u_{sR} & \sim\left(3,\,1,\,2/3\right),\quad d_{tR}\sim\left(3,\,1,\,-1/3\right),
\end{align}
with $a$ in the same range as in the previous case; $s=1,\dots,4$
and $t=1,\dots,5$.

Regarding the Yukawa Lagrangian, we can write it as follows
\begin{equation}
\mathcal{L}_{\textrm{Yuk}}=\mathcal{L}_{\textrm{Yuk}}^{\rho}+\mathcal{L}_{\textrm{Yuk}}^{\eta}+\mathcal{L}_{\textrm{Yuk}}^{\chi}\quad,\label{eq:yukawa}
\end{equation}
with
\begin{eqnarray}
\mathcal{L}_{\textrm{Yuk}}^{\rho} & = & \alpha_{t}\bar{Q}_{L}d_{tR}\rho+\alpha_{bs}\bar{Q}_{bL}u_{sR}\rho^{*}+\text{Y}_{aa'}\varepsilon_{ijk}\left(\bar{f}_{aL}\right)_{i}\left(f_{a'L}\right)_{j}^{c}\left(\rho^{*}\right)_{k}+\text{Y}'_{aa'}\bar{f}_{aL}e_{a'R}\rho\nonumber \\
 &  & +\textrm{H.c.,}\label{5}\\
\mathcal{L}_{\textrm{Yuk}}^{\eta} & = & \beta_{s}\bar{Q}_{L}u_{sR}\eta+\beta{}_{bt}\bar{Q}_{bL}d_{tR}\eta^{*}+\textrm{H.c.},\label{6}\\
\mathcal{L}_{\textrm{Yuk}}^{\chi} & = & \gamma_{s}\bar{Q}_{L}u_{sR}\chi+\gamma{}_{bt}\bar{Q}_{bL}d_{tR}\chi^{*}+\textrm{H.c.},\label{7}
\end{eqnarray}
where $a',i,j,k=1,2,3$ and with $a$, $b$, $s$, $t$ in the same
range as in the previous case. From Eqs. (\ref{5}-\ref{7}) it can
be seen that, in general, flavor-changing neutral currents (FCNCs)
can be induced because the quark fields interact with different neutral
scalar fields simultaneously. This characteristic is shared by most
of multi-Higgs models \cite{Cheng:1987}. However, some model dependent
strategies to successfully overcome that problem have been proposed
\cite{Cheng:1987,Langacker:1988,Montero:1993}. Among those, we can
mention, for instance, choosing an appropriate direction in the VEV
space, resorting to heavy scalars and/or small mixing angles in the
quark and the scalar sectors, and considering adequate Yukawa coupling
matrix textures. In particular, in this model the non-SM quarks have
the same electric charge as the SM ones. That means that these can
mix with the latter ones and hence also induce FCNC. Despite that,
this kind of FCNC is suppressed when the VEV which mainly controls
the exotic quark masses is taken much larger than the electroweak
mass scale \cite{Langacker:1988}. Also, FCNC occurs in models which
have an extra neutral vector boson. These can be handled in a similar
way. See, for example, \cite{Benavides:2009,Machado:2013}. Finally,
we remark that from Eq. (\ref{5}) it is clear that the lepton sector
of the model is not afflicted by FCNC.

We also have that the most general scalar potential consistent with
gauge invariance and renormalizability is given by
\begin{eqnarray}
V\left(\eta,\rho,\chi\right) & = & V_{\mathbb{Z}_{2}}\left(\eta,\rho,\chi\right)+V_{\cancel{\mathbb{Z}_{2}}}\left(\eta,\rho,\chi\right);\label{scalarpotential}\\
 & \textrm{with }\nonumber \\
V_{\mathbb{Z}_{2}}\left(\eta,\rho,\chi\right) & = & -\mu_{1}^{2}\eta^{\dagger}\eta-\mu_{2}^{2}\rho^{\dagger}\rho-\mu_{3}^{2}\chi^{\dagger}\chi\nonumber \\
 &  & +\lambda_{1}\left(\eta^{\dagger}\eta\right)^{2}+\lambda_{2}\left(\rho^{\dagger}\rho\right)^{2}+\lambda_{3}\left(\chi^{\dagger}\chi\right)^{2}+\lambda_{4}\left(\chi^{\dagger}\chi\right)\left(\eta^{\dagger}\eta\right)\nonumber \\
 &  & +\lambda_{5}\left(\chi^{\dagger}\chi\right)\left(\rho^{\dagger}\rho\right)+\lambda_{6}\left(\eta^{\dagger}\eta\right)\left(\rho^{\dagger}\rho\right)+\lambda_{7}\left(\chi^{\dagger}\eta\right)\left(\eta^{\dagger}\chi\right)\nonumber \\
 &  & +\lambda_{8}\left(\chi^{\dagger}\rho\right)\left(\rho^{\dagger}\chi\right)+\lambda_{9}\left(\eta^{\dagger}\rho\right)\left(\rho^{\dagger}\eta\right)+[\lambda_{10}\left(\chi^{\dagger}\eta\right)^{2}+\textrm{H.c.}];\\
V_{\cancel{\mathbb{Z}_{2}}}\left(\eta,\rho,\chi\right) & = & -\mu_{4}^{2}\chi^{\dagger}\eta\nonumber \\
 &  & +\lambda_{11}\left(\chi^{\dagger}\eta\right)\left(\eta^{\dagger}\eta\right)+\lambda_{12}\left(\chi^{\dagger}\eta\right)\left(\chi^{\dagger}\chi\right)+\lambda_{13}\left(\chi^{\dagger}\eta\right)\left(\rho^{\dagger}\rho\right)\nonumber \\
 &  & +\lambda_{14}\left(\chi^{\dagger}\rho\right)\left(\rho^{\dagger}\eta\right)+\frac{f}{\sqrt{2}}\epsilon_{ijk}\eta_{i}\rho_{j}\chi_{k}+\textrm{H.c.}
\end{eqnarray}
We have divided the total scalar potential $V\left(\eta,\rho,\chi\right)$
in two pieces, $V_{\mathbb{Z}_{2}}\left(\eta,\rho,\chi\right)$ and
$V_{\cancel{\mathbb{Z}2}}\left(\eta,\rho,\chi\right)$, for future
convenience. The first part is invariant under a $\ZZ_2$ discrete
symmetry ($\chi\rightarrow-\chi$, $u_{4R}\rightarrow-u{}_{4R}$,
$\,d{}_{\left(4,5\right)R}\rightarrow-d{}_{\left(4,5\right)R}$) in
contrast to the second one which is not. The model with such $\ZZ_2$
symmetry will be studied in detail in the next section.

The minimal vacuum structure in order to give masses for all the fields
in the model is
\[
\left\langle \rho\right\rangle =\frac{1}{\sqrt{2}}\left(\begin{array}{ccc}
0 & v_{\rho_{2}} & 0\end{array}\right)^{\textrm{T}},\,\,\left\langle \eta\right\rangle =\frac{1}{\sqrt{2}}\left(\begin{array}{ccc}
v_{\eta_{1}} & 0 & 0\end{array}\right)^{\textrm{T}},\,\,\left\langle \chi\right\rangle =\frac{1}{\sqrt{2}}\left(\begin{array}{ccc}
0 & 0 & v_{\chi_{3}}\end{array}\right)^{\textrm{T}}.
\]
Specifically, the symmetry breaking pattern is done in two stages.
First, when $\chi$ gains a VEV, $\left\langle \chi\right\rangle $,
the exotic quarks gain masses and the symmetry, $\textrm{SU}\left(3\right)_{C}\otimes\textrm{SU}\left(3\right)_{L}\otimes\textrm{U}\left(1\right)_{N}$,
is broken down to $\textrm{SU}\left(3\right)_{C}\otimes\textrm{SU}\left(2\right)_{L}\otimes\textrm{U}\left(1\right)_{Y}$.
After that, the VEV $\left\langle \rho\right\rangle $ gives \textcolor{black}{mass}
to the three charged leptons and to two of the neu\textcolor{black}{trinos
\cite{Z2_331_3,neutrino_masses,neutrino_masses_2}. Also, two up-type
quarks and one down-type quark gain masses from }$\left\langle \rho\right\rangle $.
Finally, $\left\langle \eta\right\rangle $ gives mass for the remaining
quarks. In this last stage ($\left\langle \rho\right\rangle $ and
$\left\langle \eta\right\rangle $ different from zero) the symmetry,
$\textrm{SU}\left(3\right)_{C}\otimes\textrm{SU}\left(2\right)_{L}\otimes\textrm{U}\left(1\right)_{Y}$,
is broken down to $\textrm{U}\left(1\right)_{Q}$. Note that, although
the $\eta$ and $\chi$ scalar triplets are in the same representation
of the gauge symmetries, we have defined, without loss of generality,
the $\chi$ triplet as the one responsible for the first symmetry
breaking stage. In other words, it is assumed that $\left\langle \chi\right\rangle >\left\langle \rho\right\rangle ,\,\left\langle \eta\right\rangle $.

\section{Model with $\ZZ_2$ symmetry\label{Z2symmetry}\label{sec:Model-with-Z2}}

It is a common practice to impose a discrete $\ZZ_2$ symmetry given
by: $\chi\rightarrow-\chi$, $u_{4R}\rightarrow-u{}_{4R}$, $\,d{}_{\left(4,5\right)R}\rightarrow-d{}_{\left(4,5\right)R}$
\textcolor{black}{and all}\textcolor{green}{{} }\textcolor{black}{the
other fields being even under }$\ZZ_2$\textcolor{black}{{} \cite{Z2_331_3,Z2_331,Dias_Pires_daSilva_331,Z2_331_2,Mizukoshi_Pires_Queiroz_daSilva_331,Bruce_Juan_331_PQ,331_Farinaldo,neutrino_masses}.
}This symmetry brings simplicity to the model allowing, for instance,
to interpret the $\chi$ scalar as the responsible for the first step
in the symmetry breaking pattern and in some sense, to mitigate the
FCNC issues. Besides that, it also largely simplifies the scalar potential.
In this scenario, the Yukawa Lagrangian interactions given in Eqs.
(\ref{5}-\ref{7}) are slightly modified to
\begin{eqnarray}
\mathcal{L}_{\textrm{Yuk}}^{\rho} & = & \alpha_{a}\bar{Q}_{L}d_{aR}\rho+\alpha_{ba}\bar{Q}_{bL}u_{aR}\rho^{*}+\text{Y}_{aa'}\varepsilon_{ijk}\left(\bar{f}_{aL}\right)_{i}\left(f_{bL}\right)_{j}^{c}\left(\rho^{*}\right)_{k}+\text{Y}'_{aa'}\bar{f}_{aL}e_{a'R}\rho+\nonumber \\
 &  & \textrm{H.c.,}\label{10}\\
\mathcal{L}_{\textrm{Yuk}}^{\eta} & = & \beta_{a}\bar{Q}_{L}u_{aR}\eta+\beta{}_{ba}\bar{Q}_{bL}d_{aR}\eta^{*}+\textrm{H.c.},\label{11}\\
\mathcal{L}_{\textrm{Yuk}}^{\chi} & = & \gamma_{4}\bar{Q}_{L}u_{4R}\chi+\gamma{}_{b\left(b+2\right)}\bar{Q}_{bL}d_{\left(b+2\right)R}\chi^{*}+\textrm{H.c.}\,.\label{12}
\end{eqnarray}

Furthermore, the $\ZZ_2$ symmetry forbids the terms in $V_{\cancel{\mathbb{Z}_{2}}}\left(\eta,\rho,\chi\right)$
to appear in the scalar potential. It implies that the model has actually
a larger symmetry group. Specifically, we show in Table \ref{table1}
all the U$(1)$ symmetries that the model really has (the global and
local ones). Note there are two extra global symmetries, the baryonic
one, U$(1)_{B}$, that remains unbroken, and the U$(1)_{\textrm{PQ}}$.
The last one being a Peccei-Quinn like symmetry because it is anomalous
in the color group. We also remark that these are symmetries of the
entire Lagrangian.
\begin{table}
\begin{tabular}{llccccccr||@{\extracolsep{0pt}.}lr||@{\extracolsep{0pt}.}lccccccccc}
 &  &  &  &  &  &  &  & \multicolumn{2}{c}{} & \multicolumn{2}{c}{} &  &  &  &  &  &  &  &  & \tabularnewline
\hline 
\hline 
 &  & $Q_{L}$  & $\;$$\;$  & $Q_{iL}$  & $\;$$\;$  & $\left(u_{aR},\,u_{4R}\right)$  & $\;$$\;$  & \multicolumn{2}{c}{$\left(d_{aR},d_{\left(4,5\right)R}\right)$} & \multicolumn{2}{c}{$\;$$\;$} & $f_{aL}$  & $\;$$\;$  & $e_{aR}$  & $\;$$\;$  & $\eta$  & $\;$$\;$  & $\rho$  & $\;$$\;$  & $\chi$\tabularnewline
\hline 
U$(1)_{N}$  &  & 1/3  & $ $  & 0  &  & 2/3  &  & \multicolumn{2}{c}{-1/3} & \multicolumn{2}{c}{} & -1/3  &  & -1  &  & -1/3  &  & 2/3  &  & -1/3\tabularnewline
U$(1)_{B}$  &  & 1/3  & $ $  & 1/3  &  & 1/3  &  & \multicolumn{2}{c}{1/3} & \multicolumn{2}{c}{} & 0  &  & 0  &  & 0  &  & 0  &  & 0\tabularnewline
U$(1)_{\textrm{PQ}}$  &  & 1  & $ $  & -1  &  & 0  &  & \multicolumn{2}{c}{0} & \multicolumn{2}{c}{} & -1/2  &  & -3/2  &  & 1  &  & 1  &  & 1\tabularnewline
\hline 
\end{tabular}

\caption{The U$(1)$ symmetries in the model when the $\ZZ_{2}$ discrete symmetry
is considered.\label{table1}}
\end{table}

In the minimal case, when only three VEVs - $v_{\eta_{1}},v_{\rho_{2}}$
and $v_{\chi_{3}}$ - are different from zero, the scalar sector has
a Nambu-Goldstone (NG) boson, $J$, in the physical spectrum. It is
given by 
\begin{equation}
J=\frac{1}{N_{J}}\left(\frac{v_{\eta_{1}}v_{\chi_{3}}}{v_{\rho_{2}}}\textrm{Im}\,\rho_{2}^{0}+v_{\chi_{3}}\textrm{Im}\,\eta_{1}^{0}+v_{\eta_{1}}\textrm{Im}\,\chi_{3}^{0}\right),\label{J1}
\end{equation}
with $N_{J}\equiv\left(v_{\eta_{1}}^{2}v_{\chi_{3}}^{2}v_{\rho_{2}}^{-2}+v_{\eta_{1}}^{2}+v_{\chi_{3}}^{2}\right)^{1/2}$.
We emphasize that $J$ in Eq. (\ref{J1}) is orthogonal to the NG
bosons which are absorbed by the gauge vector bosons, as it should
be. We have followed the method described in Refs. \cite{Srednicki1985689,Srednicki_book_2007}
to accomplish that. 

From the explicit form of $J$ in Eq. (\ref{J1}), and the Lagrangian
in Eq. (\ref{10}), it is straightforward to calculate the coupling
of $J$ with the electron and positron, $g_{e\bar{e}J}$. It is explicitly
given by 
\begin{eqnarray}
g_{e\bar{e}J} & = & \frac{\sqrt{2}m_{e}v_{\eta_{1}}v_{\chi_{3}}}{N_{J}v_{\rho_{2}}^{2}},\label{geej}
\end{eqnarray}
where $m_{e}$ is the electron mass. The existence of this coupling
opens a new channel of energy loss in stars through the Compton-type
process $\gamma+e^{-}\rightarrow e^{-}+J$. From the evolution of
red-giant stars we have that \cite{Tabulation_astrophysical_1987,RedGiant_Raffelt_1995,eeJ_old1,eeJ_old2,eeJ_old3}
\begin{eqnarray}
\left|g_{e\bar{e}J}\right| & \lesssim & g_{\textrm{max}}\equiv10^{-13}.\label{gbound}
\end{eqnarray}
In order to impose a bound on the VEVs, we bring another piece of
information. The mass of the $W^{\pm}$ bosons is
\begin{eqnarray}
M_{W^{\pm}}^{2} & = & \frac{g_{L}^{2}}{4}\left(v_{\eta_{1}}^{2}+v_{\rho_{2}}^{2}\right),\label{mw}
\end{eqnarray}
where $g_{L}$ is the gauge coupling constant of the $\textrm{SU}\left(3\right)_{L}$
group. Thus, $v_{\eta_{1}}^{2}+v_{\rho_{2}}^{2}=v_{\textrm{SM}}^{2}\simeq246^{2}$
GeV$^{2}$ in order to obtain the SM $W^{\pm}$ mass \cite{Olive:2014PDG}.
From Eqs. (\ref{geej}-\ref{mw}) the following upper bound on $v_{\chi_{3}}$
can be found
\begin{equation}
v_{\chi_{3}}\leq v_{\chi\textrm{max}}(v_{\rho_{2}})\equiv v_{\rho_{2}}\left[2g_{\textrm{max}}^{-2}m_{e}^{2}/v_{\rho_{2}}^{2}-1/\left(1-v_{\rho_{2}}^{2}/v_{\textrm{SM}}^{2}\right)\right]^{-1/2}.\label{vchimax}
\end{equation}
Thus, the upper bound on $v_{\chi_{3}}$ is a function of $v_{\rho_{2}}$.
Because of Eq. (\ref{vchimax}), we can estimate that the largest
value that $v_{\chi_{3}}$ can take is $v_{\chi\textrm{max}}\left(v_{\rho_{2}}\rightarrow v_{\text{SM}}^{-}\right)\simeq11.5$
keV. However, it is in contradiction with the general assumption in
this model which claims that $\left\langle \chi\right\rangle >\left\langle \rho\right\rangle ,\,\left\langle \eta\right\rangle $.
In addition to that, there is a fact that rules out this scenario
with the $\ZZ_2$ symmetry. In order to understand it, notice that
the NG boson in Eq. (\ref{J1}), that results from the breaking of
the symmetries, is actually an axion because of the U$(1)_{\textrm{PQ}}$
symmetry in Table \ref{table1}. Moreover, the decay constant $f_{a}$
for the axion is, in this case, given by
\begin{equation}
{\color{black}f_{a}=N_{J}.}
\end{equation}
From this and the upper bound in Eq. (\ref{vchimax}) we can find
an upper bound on $f_{a}$ as follows
\begin{equation}
f_{a}\leq\,f_{a\,\textrm{max}}(v_{\rho_{2}})\equiv v_{\text{SM}}\left[1-\frac{v_{\rho_{2}}^{2}}{v_{\text{SM}}^{2}}+\frac{v_{\chi\textrm{max}}^{2}(v_{\rho_{2}})}{v_{\rho_{2}}^{2}}\right]^{1/2}.
\end{equation}
From the equation above, we can see that $v_{\chi\textrm{max}}(v_{\text{SM}})\approx11.5\text{ keV}\leq f_{a}\leq v_{\text{SM}}$.
However, an axion with this small decay constant was ruled out long
ago \cite{axion_limits_Preskill,axions_limits_Peccei}.

In the symmetry breaking patterns with more than three VEVs different
from zero, i.e., $\left(v_{\eta_{1}},v_{\eta_{3}},v_{\rho_{2}},v_{\chi_{3}}\right)$,
$\left(v_{\eta_{1}},v_{\rho_{2}},v_{\chi_{1}},v_{\chi_{3}}\right)$,
and $\left(v_{\eta_{1}},v_{\eta_{3}},v_{\rho_{2}},v_{\chi_{1}},v_{\chi_{3}}\right)$,
the situation is even worse. Although the exact form of the $CP$-even
and $CP$-odd scalars present in the physical spectra depends on the
particular breaking pattern of the symmetry, there are some general
features that we can mention. In all cases, there are two physical
NG bosons, $J_{I}$ and $J_{R}$, the first in the $CP$-odd sector,
and the second in the $CP$-even sector. One of them couples with
the electron and positron, i.e., $g_{e\bar{e}J_{I}}\neq0$, which
imposes upper bounds on the VEVs as in the previous case. Moreover,
the existence of $J_{R}$ in the physical spectrum brings an extra
difficulty that rules out the model once and for all when that $\ZZ_{2}$
symmetry is considered. This difficulty comes from the invisible decay
width of the $Z$ gauge boson which receives at least one new contribution,
$Z\rightarrow J_{R}J_{I}$, and there is no more room for this \cite{Olive:2014PDG}.
Therefore, we can conclude that the model invariant under the $\ZZ_{2}$
symmetry is not consistent with constraints coming from astrophysics
and particle physics. 

\section{Model with soft $\ZZ_2$-breaking terms\label{SZ2symmetry}\label{sec:Model-with-soft}}

There is another possibility to be taken into account when the $\ZZ_2$
discrete symmetry is considered. This is to introduce some soft $\ZZ_2$-breaking
terms in the scalar potential in order to remove the extra accidental
symmetries in the Lagrangian, and at the same time leave the model
as simple as possible. With that in mind, we are going to explore
two terms, $\mu_{4}^{2}\chi^{\dagger}\eta$ and $\frac{f}{\sqrt{2}}\epsilon_{ijk}\eta_{i}\rho_{j}\chi_{k}$,
that can in principle accomplish that. Let us consider both cases
separately by starting from the $\mu_{4}^{2}\chi^{\dagger}\eta$ term. 

\subsection{$\mu_{4}^{2}\,\chi^{\dagger}\eta$ term}

First, we consider the $\mu_{4}^{2}\,\chi^{\dagger}\eta$ soft-breaking
term together with the scalar potential allowed by the $\mathbb{Z}_{2}$
symmetry. Although this term does not remove any extra symmetry in
Table \ref{table1}, its presence slightly changes the form of the
NG boson in comparison to the case of the model without it. In some
sense that justifies our next analysis. In order to extract conclusions
we obtain the physical NG boson for each symmetry breaking pattern.
The NG boson looks in each case as follows
\begin{eqnarray}
v_{\eta_{1}},\,v_{\rho_{2}},\,v_{\chi_{3}} & \rightarrow & J_{1}=N_{J_{1}}^{-1}\left(v_{\chi_{3}}\textrm{Im}\,\eta_{1}^{0}+\frac{v_{\eta_{1}}v_{\chi_{3}}}{v_{\rho_{2}}}\textrm{Im}\,\rho_{2}^{0}+v_{\eta_{1}}\textrm{Im}\,\chi_{3}^{0}\right),\label{eq32}\\
v_{\eta_{1}},\,v_{\eta_{3}},\,v_{\rho_{2}},\,v_{\chi_{3}} & \rightarrow & J_{2}=N_{J_{2}}^{-1}\left(v_{\chi_{3}}\textrm{Im}\,\eta_{1}^{0}+\frac{v_{\eta_{1}}v_{\chi_{3}}}{v_{\rho_{2}}}\textrm{Im}\,\rho_{2}^{0}+v_{\eta_{1}}\textrm{Im}\,\chi_{3}^{0}\right.\nonumber \\
 &  & \quad\quad\quad\quad\quad\left.-v_{\eta_{3}}\textrm{Im}\,\chi_{1}^{0}\right),\\
v_{\eta_{1}},\,v_{\rho_{2}},\,v_{\chi_{1}},\,v_{\chi_{3}} & \rightarrow & J_{3}=N_{J_{3}}^{-1}\left(v_{\chi_{3}}\textrm{Im}\,\eta_{1}^{0}+\frac{v_{\eta_{1}}v_{\chi_{3}}}{v_{\rho_{2}}}\textrm{Im}\,\rho_{2}^{0}+v_{\eta_{1}}\textrm{Im}\,\chi_{3}^{0}\right.\nonumber \\
 &  & \quad\quad\quad\quad\quad\left.-v_{\chi_{1}}\textrm{Im}\,\eta_{3}^{0}\right),\label{eq33}\\
v_{\eta_{1}},\,v_{\eta_{3}},\,v_{\rho_{2}},\,v_{\chi_{1}},\,v_{\chi_{3}} & \rightarrow & J_{4}=N_{J_{4}}^{-1}\left(v_{\chi_{3}}\textrm{Im}\,\eta_{1}^{0}-\frac{\left(v_{\eta_{3}}v_{\chi_{1}}-v_{\eta_{1}}v_{\chi_{3}}\right)}{v_{\rho_{2}}}\textrm{Im}\,\rho_{2}^{0}+v_{\eta_{1}}\textrm{Im}\,\chi_{3}^{0}\right.\nonumber \\
 &  & \quad\quad\quad\quad\quad\left.-v_{\chi_{1}}\textrm{Im}\,\eta_{3}^{0}-v_{\eta_{3}}\textrm{Im}\,\chi_{1}^{0}\right),\label{eq34}
\end{eqnarray}
where $N_{J_{i}}$ (normalization factors) are given by $N_{J_{i}}\equiv\left(C_{\eta_{1}}^{2}+C_{\eta_{3}}^{2}+C_{\rho_{2}}^{2}+C_{\chi_{1}}^{2}+C_{\chi_{3}}^{2}\right)^{1/2}$,
and we also have defined $J_{i}\equiv N_{J_{i}}^{-1}\left[C_{\rho_{2}}\textrm{Im}\,\rho_{2}^{0}+C_{\eta_{1}}\textrm{Im}\,\eta_{1}^{0}+C_{\eta_{3}}\textrm{Im}\,\eta_{3}^{0}+C_{\chi_{1}}\textrm{Im}\,\chi_{1}^{0}+C_{\chi_{3}}\textrm{Im}\,\chi_{3}^{0}\right]$. 

One can note that the component $\textrm{Im}\,\rho_{2}^{0}$ in $J$
accounts for a non-zero value of $g_{e\bar{e}J}$, which is generically
given by $g_{e\bar{e}J_{i}}=\frac{\sqrt{2}m_{e}}{v_{\rho_{2}}}N_{J_{i}}^{-1}C_{\rho_{2}}$.
The case $(v_{\eta_{1}},\,v_{\rho_{2}},\,v_{\chi_{3}})$ is ruled
out following discussion after Eq. (\ref{vchimax}) since we exactly
recover the case of Eqs. (\ref{J1}) and (\ref{geej}). For the cases
with more than three non-vanishing VEVs, one has to deal with more
variables which makes these three scenarios - $J_{2}$, $J_{3}$ and
$J_{4}$ - not easy to exclude using the same strategy followed in
Section \ref{Z2symmetry}. The best we can do with that strategy is
set an upper bound on $v_{\chi_{3}}$. For the case of four VEVs we
manage to find $v_{\chi_{3}}\lesssim355\textrm{ GeV}$. The case of
the five VEVs is more intricate although a similar bound (although
not general) is possible to get. Roughly speaking, these bounds comes
from the simultaneous application of $g_{e\bar{e}J}$, $\rho=M_{W}^{2}/\left(M_{Z}^{2}\textrm{cos}^{2}\theta_{W}\right)$
bounds ($M_{W}$, $M_{Z}$ and $\theta_{W}$ are the $W$ boson mass,
$Z$ boson mass and the weak mixing angle, respectively \cite{Olive:2014PDG}). 

There is another immediate consequence arising from the $J$ form.
Following Eq. (\ref{eq:yukawa}), and taking into account the expressions
of the $J$ NG bosons in Eqs. (\ref{eq32}-\ref{eq34}), we are led
to the conclusion that they are going to interact inevitably with
the quarks of the model. It implies that the $g_{nnJ}$ coupling,
the interaction of two nucleons with $J$, is different from zero
in this case. Nevertheless, it must be smaller than $10^{-12}$, i.e.,
$\left|g_{nnJ}\right|\lesssim g_{nnJ}^{\text{max}}\equiv10^{-12}$
\cite{Tabulation_astrophysical_1987,astrophysical_2008,astro_1989,astro_1990,astrophysical_1996,neutron_stars_old1,neutron_stars_old2}.
We have checked that this bound eventually imposes strong constraints
on the VEVs and the Yukawa couplings of the model. In general, we
have seen that some parameters are forced to have an $\mathcal{O}\left(10^{-15}\right)$
tuning in order for the limit on $g_{nnJ}$ to be respected. However,
we are not going into the details because there is a different line
of reasoning that unquestionably rules out this scenario with just
the $\mu_{4}^{2}\,\chi^{\dagger}\eta$ term. It consists on the observation
that the U$(1)_{\textrm{PQ}}$ which generates the NG boson (in this
case it is an axion) is actually broken in the electroweak scale.
In order to see that, note that the $3\text{U}(1)_{N}+\text{U}(1)_{\textrm{PQ}}$
subgroup is not spontaneously broken by any $\eta_{i}$ or $\chi_{i}$
VEVs. Thus, the responsible for breaking that subgroup is the $\rho$
VEV which is upper-bounded by the electroweak scale ($V_{\text{SM}}\simeq246$
GeV) since it is mainly responsible for giving mass to all SM leptons
in this model. Therefore, the NG boson (axion) would be visible and
thus already ruled out. All in all, the soft $\mu_{4}^{2}\,\chi^{\dagger}\eta$
breaking term is not capable to get rid of the issues arising from
the presence of a NG boson in this model.

\subsection{$\frac{f}{\sqrt{2}}\epsilon_{ijk}\eta_{i}\rho_{j}\chi_{k}$ term }

From Table \ref{table1}, we can see that the $\frac{f}{\sqrt{2}}\epsilon_{ijk}\eta_{i}\rho_{j}\chi_{k}$
term removes the U$(1)_{\textrm{PQ}}$ symmetry. As a result there
are no physical NG bosons, fact which leaves the model safe regarding
the appearance of these massless particles. Therefore, we conclude
that this model can be considered with the $\ZZ_2$ symmetry provided
the soft $\frac{f}{\sqrt{2}}\epsilon_{ijk}\eta_{i}\rho_{j}\chi_{k}$
term is included in the scalar potential.

\section{Conclusions\label{conclusions}}

The presence of physical NG bosons can play a potentially important
role in constraining the parameters of a model, and in some cases
excludes it. The reason is that any light neutral particle can interact
with matter providing an important stellar energy-loss mechanism.
With this motivation, we have constrained an appealing version of
the $3-3-1$ model, taking into account the presence of physical NG
bosons, which are inconsistent with both astrophysical and particle
physics well established results. Specifically, we have considered
the $3-3-1$ model with right-handed neutrinos when a $\mathbb{Z}_{2}$
symmetry is imposed. This scenario is conventionally considered in
the literature because it greatly simplifies the model. However, that
discrete symmetry brings as a consequence the introduction of an extra
global symmetry, U$(1)_{\textrm{PQ}}$, which when spontaneously broken
by the $v_{\eta_{1}},\,v_{\rho_{2}},\,v_{\chi_{3}}$ VEVs, introduces
an axion in the model. The issue is that imposing simultaneously the
bounds on $g_{e\bar{e}J}$ and $M_{W}$ we found bounds on the decay
coupling constant $f_{a}$, specifically, $11.5\text{ keV}\leq f_{a}\leq v_{\text{SM}}$.
It is inconsistent with experiments looking for light scalars (or
pseudo-scalars) since that window for $f_{a}$ makes the axion visible.
In addition to that, we have considered all the other possibilities
for the available VEVs in the model, showing that these are also excluded.
The reason is that the appearance of physical NG bosons contributing
to the invisible decay width of the $Z$ gauge boson ($Z\rightarrow J_{R}J_{I}$)
is in conflict with experimental data. All of these scenarios are
then ruled out. 

We also studied in detail the soft $\mathbb{Z}_{2}$-symmetry breaking
case. The two terms allowed by the gauge symmetries are $\mu_{4}^{2}\,\chi^{\dagger}\eta$
and $\frac{f}{\sqrt{2}}\epsilon_{ijk}\eta_{i}\rho_{j}\chi_{k}$. In
the first case, we found that although the NG bosons have slightly
different forms, there are also problems that exclude all of those
scenarios. The main reason is that a $3\text{U}(1)_{N}+\text{U}(1)_{\textrm{PQ}}$
subgroup remains unbroken after the first step of the spontaneous
symmetry breaking. Thus, the decay coupling constant is of the order
of the electroweak scale, $v_{\text{SM}}$, implying the same issues
as the previous case. Finally, we show that the model can be considered
consistent with the $\mathbb{Z}_{2}$ symmetry when the $\frac{f}{\sqrt{2}}\epsilon_{ijk}\eta_{i}\rho_{j}\chi_{k}$
term is included in the scalar potential. That term breaks the $\mathbb{Z}_{2}$
symmetry softly and removes the extra global symmetry. Thus, no physical
NG boson appears at all and the model is safe concerning this issue.
\begin{acknowledgments}
B. L. S. V. would like to thank Coordenação de Aperfeiçoamento de
Pessoal de Nível Superior (CAPES), Brazil, for financial support.
E. R. S. would like to thank Conselho Nacional de Desenvolvimento
Científico e Tecnológico (CNPq), Brazil, for financial support and
Bonn University for kind hospitality.
\end{acknowledgments}

\bibliographystyle{unsrt}
\bibliography{references}

\end{document}